# THE UWB SOLUTION FOR MULTIMEDIA TRAFFIC IN WIRELESS SENSOR NETWORKS

A.A.BOUDHIR, M. BOUHORMA, M. BEN AHMED and ELBRAK SAID
LIST Laboratory, Faculty Of Sciences and Techniques, Tangier Morocco
CBAR, Al Yamamah university, Kingdom of Saudi Arabia
hakim.anouar@ieee.org , bouhorma@gmail.com, med.benahmed@gmail.com,elbraks@gmail.com

## ABSTRACT

*Several researches are focused on the QoS (Quality of Service) and Energy consumption in wireless Multimedia Sensor Networks. Those research projects invest in theory and practice in order to extend the spectrum of use of norms, standards and technologies which are emerged in wireless communications. The performance of these technologies is strongly related to domains of use and limitations of their characteristics. In this paper, we give a comparison of ZigBee technology, most widely used in sensor networks, and UWB (Ultra Wide Band) which presents itself as competitor that present in these work better results for audiovisual applications with medium-range and high throughput.*

## 1. INTRODUCTION

The establishment of wireless networks in various domains knew an amazing success. In front of this progress several researches are followed to enlarge the range of its use. Networks of wireless sensors are a particular Ad hoc network, integrated with an active applications allowing control, surveillance and help to decision.

The introduction of wireless communication is dramatically changing our lives. The ability to communicate anytime anywhere increases our quality of lives and improves our business productivity. The recent technological developments that allow us to execute bandwidth-hungry multimedia applications over the wireless media add new dimensions to our ability to Communicate. Various technologies appeared in sensors networks and assure the communication differently, this difference comes especially in the given quality of service and solutions given to constraints.

In this document we study the quality of service of technologies Zigbee and UWB (Ultra Wide Band) as well as the consumption of energy for a multimedia flux using the simulator NS2.This to solicit the adapted technology to the transmission of Multimedia flux in WMSN(Wireless Multimedia Sensor Network).

## 2. RELATED WORKS

Several papers focused their works on WMNs treating an enormous topics related to energy efficiency, QoS, routing protocols on MAC and physical layers. Karapistoli *et al* [1] identify the cross-layer dependencies between the specified physical layer and the higher layers of the communication. In [2] authors raise a ranging method of localization technique in WSN based on ultra-wideband (UWB) communication technology. Melodia *et al* [3], present a cross-layer communication architecture based on the time-hopping impulse radio ultra wide band technology to deliver QoS to heterogeneous applications in WMSNs, by leveraging and controlling interactions among different layers of the protocol stack according to applications requirements. Berthe *et al* [4], propose a WSN simulation architecture based on the IR-UWB technique. At the PHY layer, they take into account the pulse collision by dealing with the pulse propagation delay. They also modelled MAC protocols specific to IRUWB,for WSN applications and propose a generic and reusable sensor and sensing channel model. Most of the WSN application performances can be evaluated thanks to this simulation architecture.

## 3. WIRELESS SENSOR NETWORKS

Wireless sensor networks (WSNs) contain hundreds or thousands of sensor nodes equipped with sensing, computing and communication abilities. Each node has the ability to sense elements of its environment, perform simple computations, and communicate among its peers or directly to an external base station or sink (figure.1). The node or mote that needs to operate for a long time on a tiny battery is composed of a processor, a memory, a transmitter/ receiver radio, an embedded system composed of a unit of sensing and a battery (figure2). This component can be in sleep mode or listen only to the traffic. The unit of transmission is the unit which uses most energy compared to others units constituting a sensor [05].

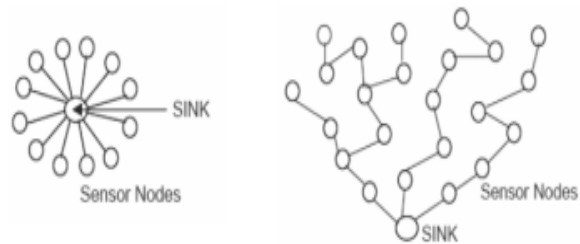

Figure.1: Architecture of Wireless Sensor Network

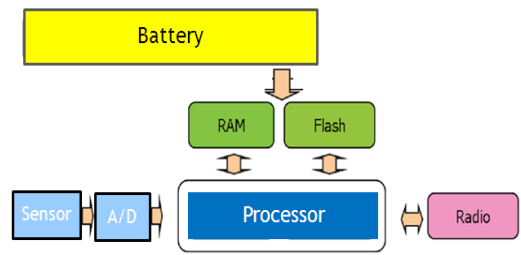

Figure.2: Architecture of Node

## 4. MULTIMEDIA WIRELESS SENSOR NETWORKS

The technology development in electronics contributed to the availability of miniaturized materials with low cost such as CMOS cameras and microphones which helped more the development of wireless multimedia sensor networks (WMSN), devices that are able to retrieve multimedia content such as the ubiquitous audio and video, still images, and data from environmental sensors.

Wireless multimedia sensor networks will not only reinforce the networks of sensors such as monitoring, home automation and environmental monitoring, but will also enable several new applications like monitoring networks multimedia, storage of potentially activities, control systems traffic, medical surveillance environmental monitoring, location services, industrial process control…

## 5. THE QUALITY OF SERVICE

In telecommunication networks, the goal of QoS is to reach a better behavior of communication for the content which must be properly routed, and network resources are used optimally [06]. Generally, researches on QoS in wireless networks in several key areas; models of QoS differentiation at the MAC layer (Medium Access Control) protocols for signaling and routing with QoS. The need of QoS can be specified into measurable parameters mentioned in (3), (4) and (5):

-*End to End Delay:*
$$EED = \frac{Time\ spent\ to\ deliver\ packets}{\sum received\ packets} \quad (3)$$

-*Bandwidth:*
$$BW = packets\ size \times \frac{\sum Received\ packets}{End\ Time\ Simulation} \quad (4)$$

-*Packet delivery ratio:*
$$PDR\ (\%) = 100 \times \frac{\sum Received\ packets}{\sum sent\ packets} \quad (5)$$

## 6. TECHNOLOGIES EMERGED IN WIRELESS SENSOR COMMUNICATION

Many technologies are allowed to wireless transmission of information. Each represents a different use, according to its characteristics (transmission speed, maximum flow, Cost of infrastructure cost of equipment connected Security, Flexibility of installation and use, power consumption and autonomy …).

### 6.1. ZigBee Technology

ZigBee is a specification for a suite of high level communication protocols using small, low-power digital radios based on the IEEE 802.15.4 standard for wireless personal area networks (WPANs), such as wireless headphones connecting with cell phones via short-range radio. The technology defined by the ZigBee specification is intended to be simpler and less expensive than other WPANs, such as Bluetooth. ZigBee is targeted at radio-frequency (RF) applications such as industrial control and monitoring, wireless sensor networks, asset and inventory tracking, intelligent agriculture, and security would benefit from such a network topology that require a low data rate, long battery life, and secure networking.

ZigBee builds upon the physical layer and medium access control defined in IEEE standard 802.15.4 for low-rate WPAN's. The specification goes on to complete the standard by adding four main components: network layer, application layer, ZigBee device objects (ZDO's) and manufacturer-defined application objects which allow for customization and favor total integration.

ZigBee operates in the industrial, scientific and medical (ISM) radio bands; 868 MHz in Europe, 915 MHz in the USA and Australia, and 2.4 GHz in most jurisdictions worldwide. (figure.3)[07].

## 6.2. UWB Technology

Ultra Wide Band (UWB) technology based on sending pulses of energy low power over a wide frequency band is able to communicate wirelessly as an indoor short-range high-speed communication. One of the most exciting characteristics of UWB is that its bandwidth is over 110 Mbps (up to 480 Mbps) which can satisfy most of the multimedia applications, especially in wireless sensor networks, such as audio and video delivery in home networking and it can also act as a wireless cable replacement of high speed serial bus such as USB 2.0 and IEEE 1394. UWB works via chip-based radios that modulate signals across the entire available ultra wideband spectrum, which in the US is from 3.1 to 10.6 GHz (figure.4) [09]. In [08] several standards are mentioned giving the differences among four technologies. Each one is based on an IEEE standard.

Obviously, UWB and Wi-Fi provide a higher data rate, while Bluetooth and ZigBee give a lower one. In general, the Bluetooth, UWB, and ZigBee are intended for WPAN communication (about 10m), while Wi-Fi is oriented to WLAN (about 100m). However, ZigBee can also reach 100m in some applications.

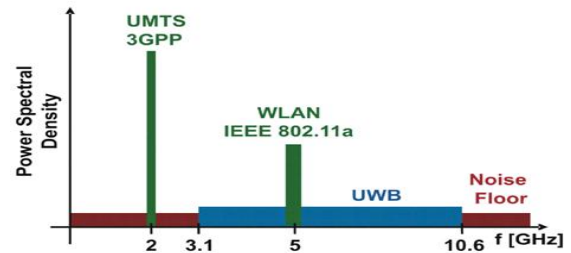

Figure.3: The three frequency band for IEEE 802.15.4 standard    Figure.4: Frequency Spectrum in UWB technology

## 7. SIMULATION AND RESULTS

### 7.1. Environment Of Simulation

The simulation tool used is the NS2 simulator dedicated to wireless networks and considered a crucial asset search.

The version of ns2-allinone-2.29 [10]used, incorporate into the architecture of the MAC layer (mac.cc / mac.h) and physical (phy.cc / phy.h) modules and standard IEEE.802.15.4 supporting radio pulses compliant to IEEE 802.15.3 UWB which adds to its MAC layer modules DCC-MAC layer (mac-ifcontrol*. (cc, h)) and physical layer (interference-phy*. (cc, h) ) by implementing the NOAH protocol that allows direct communications (unlike AODV, DSR, ...) between wireless nodes, or between base stations and mobile nodes. It can simulate scenarios where multi-hop routing is undesirable.

*Mac IFcontrol* [10]: Defines the MAC layer for UWB, functions of transmission, queue management, control packets and listening mode etc.

*Interference-phy* [10]: defines possible states (reception, transmission, listen or hang) and manages the time to listen and reception etc.

## 7.2. PARAMETERS OF SIMULATIONS

In order to evaluate the quality of service, the simulations treat a comparative metric subject to two different multicast protocols AODV [11], [12] and DSR [11],[12] for ZigBee technology and protocol NOAH for UWB parameters as mentioned below. The figure 5 list an example of architecture adopted, using node as wireless multimedia sensors and a collector of data "C" (node0 in NAM Visualisator). The simulation results are drawn from the files "tr" generated and analyzed by file "awk".

The table 2 summarizes the simulation parameters used for the ZigBee and UWB technologies:

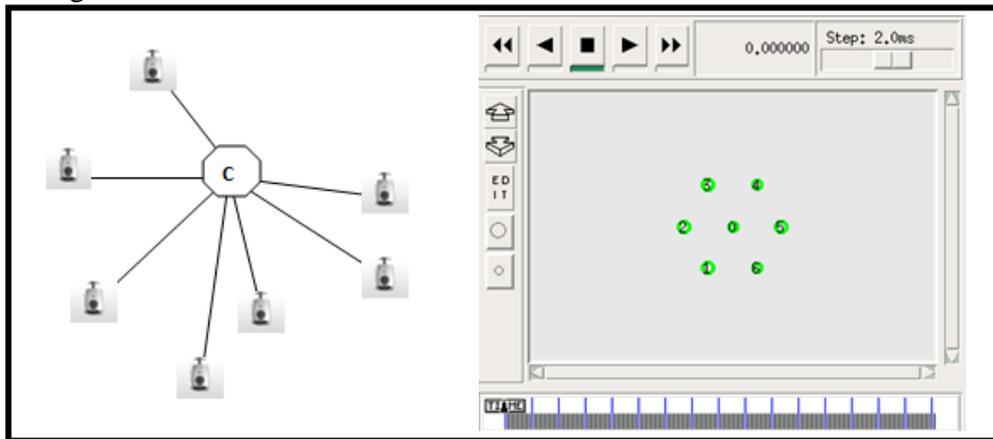

fig.5: Sample of architecture adopted in simulations

TABLE 2
Parameters used in ZigBee and UWB simulations

| Technology | ZigBee | UWB |
|---|---|---|
| Protocol | AODV / DSR | NOAH |
| Mac /phy | 802_15_4 | 802_15_3 |
| Chanel | Wireless Channel | InterferencePhy |
| Propagation | TwoRayGround | PropTarokh |
| Topology | 20*20 | 20*20 |
| Traffic | UDP/FTP | UDP/ FTP |
| Number of nodes | 7 – 25 - 101 | 7 - 25 - 101 |
| RtPower | 0.00075 w | 0.00075 w |
| TxPower | 0.00175 w | 0.00175 w |
| Initial Energy | 1000 j | 1000 j |
| Sleep Energy | 0.00005j | 0.00005j |

## 7.3. RESUALTS AND DISCUSSION

The following figures illustrate a comparison based on the results of previous simulations show the benefit of UWB in the delivery of a high rate of packets (fig7 and fig8) compared to ZigBee (fig6) with a wide bandwidth (fig9) and a delay (fig10) start to finish while consuming a minimum of Energy (fig11)for less dense networks:

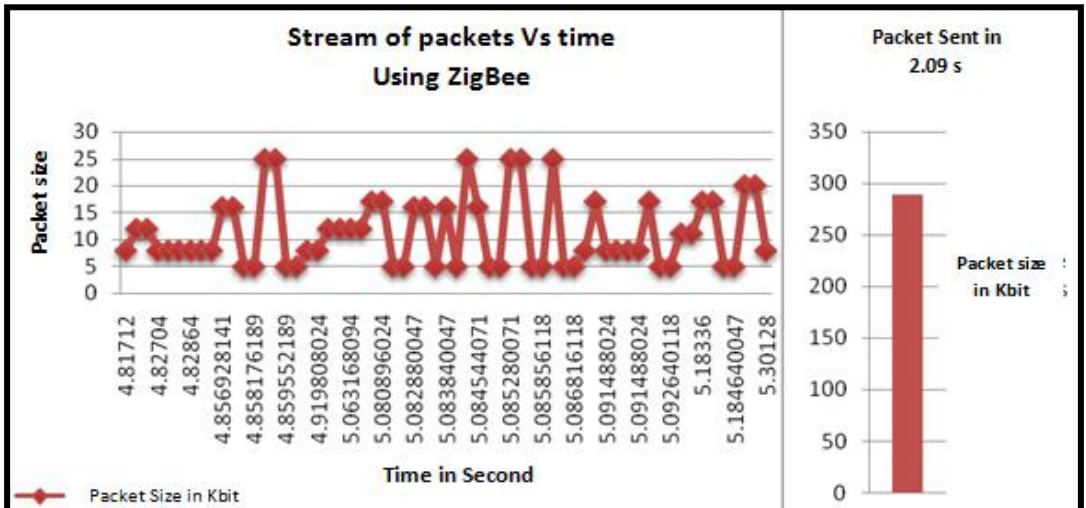
fig.6 : Stream Packets Vs Time Using ZigBee

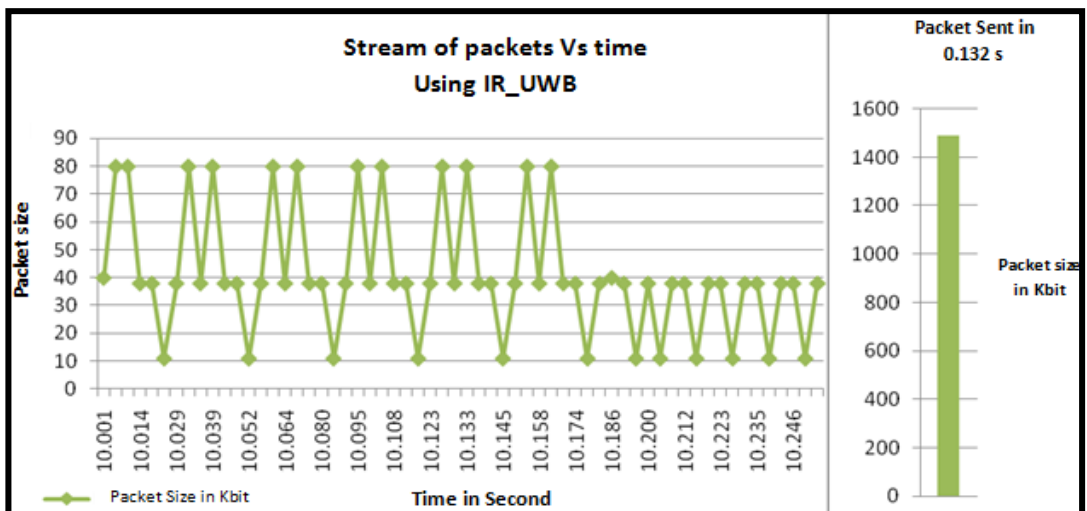
fig.7 : Stream Packets Vs Time Using impulse radion UWB

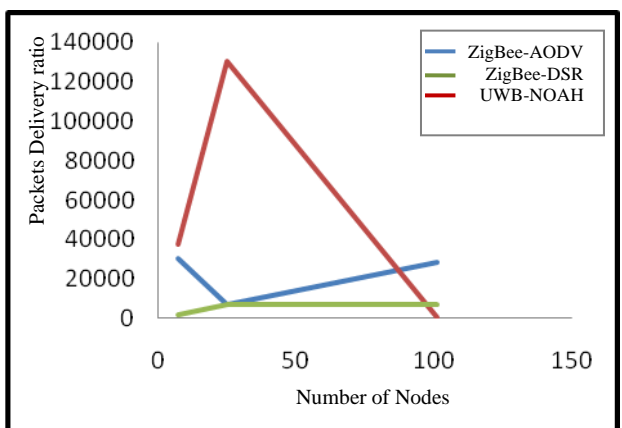
Fig.8: Packets Delivery Ratio

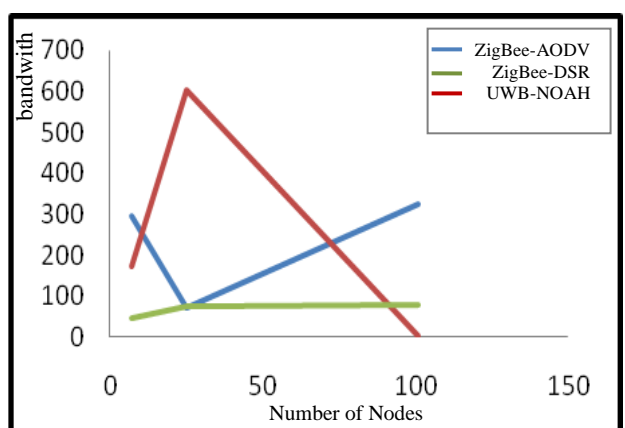
Fig.9: Bandwidth

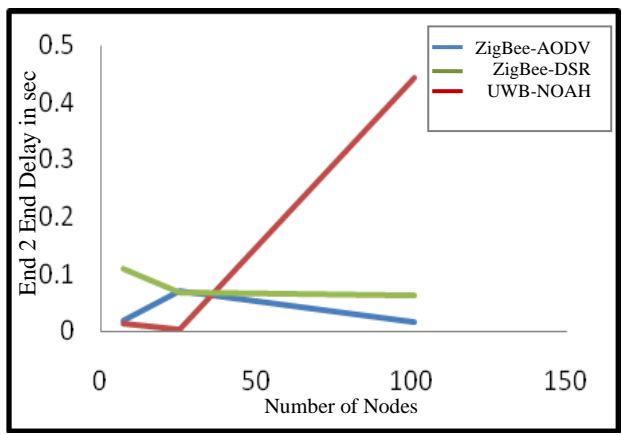
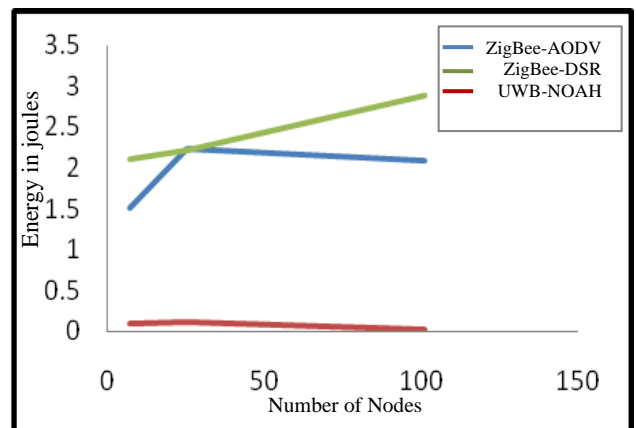

Fig.10: End to End Delay

Fig.11: Energy Consumption

With the simulation parameters mentioned, the UWB technology offered a considerable quality of service in term of end to end delay, packet delivery ratio, bandwith and energy for a sensor network with less density of network when sending media streams. Indeed, this technology has gives a good results compared to those of the ZigBee technology especially for a network average of 25 nodes which promotes greater use of these sensors for the transfer of multimedia data.

## 8. CONCLUSION

This work has explored a survey on UWB and ZigBee standars in order to compare their performance in wireless multimedia sensor network context. We have presented in this article a quality of service of standards mentionned above, using the popular metrics as a network performance tools. In particular, those metrics, already presented, were evaluated to study the quality of communication of multimedia data under the network simulator NS2.

Furthermore, we found that the performance tests conducted on the consumption of energy, packet delivery ratio, bandwidth and end-to-end delay, have shown that the UWB technology responds well performance criteria desired. Indeed, the ultra wide band (UWB) technology has the potential to enable low-power consumption, high data rate communications within tens of meters, characteristics that make it an ideal choice for WMSNs.

**BOUDHIR ANOUAR ABDELHAKIM,** born in FES, Morocco, in1979.Actualy Member of IEEE since 2009. He received the "Master of Sciences and Technologies" degree in Electrical Engineering from the university CADI AYYAD of Marrakesh and recently in 2009, the "Magister in Computer Sciences Systems and Networks" degree From the University ADBELMALEK ESSAADI of Tangier. He is currently working toward the Ph.D. degree with the Computing and Telecommunications Research Group at Abdelmalek Essaadi University. His research interests include security and quality of service in wireless sensor networks. 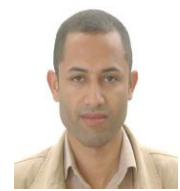

**Mohammed BOUHORMA**, Prof. Dr. received the "Master degree in Electronics" degree in 1990 from Abdelmalek Essaadi University in Tetuan, Morocco, the "DEA" degree in Electronics and Telecommunications and the "Doctorat d'Etat" degree in Telecommunications with honors, respectively, in 1991 and 1995 from the "ENSEEIHT-INPT" of Toulouse. He is a Professor of computer sciences and networks at Abdelmalek Essaadi University since 1999.He was chief of computer sciences department and he is also responsible of the Master titled Computer Sciences Systems and Computers.Prof. BOUHORMA has supervised several Ph D and Masters Theses and has been the principal investigator and the project manager for several international research projects dealing with different research topics concerned with his research interests mentioned above. 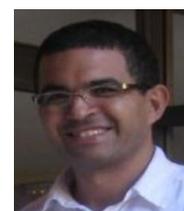

**Mohamed Ben Ahmed** was born in Al-Hoceima, Morocco, in December19, 1979. He received the "Master of Science and Technology" degree in Computer Sciences ,the "DESA" degree in Telecommunications, and the "PhD" degree in Computer Sciences and Telecommunications respectively, in 2002, 2005 and 2010 from Abdelmalek Essaadi University in Tetuan, Morocco. He is currently a Professor Assistant in Abdelmalek Essaadi University. His research interests include multi-Bands antenna and SAR calculation. 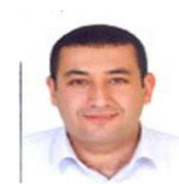

**El Brak SAID**, born in Ksar Kbir, Morocco, in1979. He received the DESA degree in Electrical Engineering from the university ABDELMALEK ESSADI of Tetouan. Currently he is working toward the Ph.D. degree with the Computing and Telecommunications Research Group at Abdelmalek Essaadi University. His research interests include Ad hoc and sensor networks